\documentclass[prl,aps,twocolumn,showpacs]{revtex4}
\usepackage{bm}
\usepackage{epsfig}
\usepackage{amsmath}

\newcommand{\bpm}{\begin{pmatrix}}
\newcommand{\epm}{\end{pmatrix}}
\newcommand{\ba}{\begin{eqnarray}}
\newcommand{\ea}{\end{eqnarray}}

\begin{document}

\title{Broken relationship between superconducting pairing interaction and electronic dispersion kinks in La$_{2-x}$Sr$_x$CuO$_4$}

\author{S. R. Park,$^{1}$ Y. Cao,$^{1}$ Q. Wang,$^{1}$ M. Fujita,$^{2}$ \\K. Yamada,$^{3}$ S.-K. Mo,$^{4}$ D. S. Dessau,$^{1}$ D. Reznik,$^{1}$}
\email[Electronic address:$~~$]{dmitry.reznik@colorado.edu}
\affiliation{$^{1}$Department of physics, University of Colorado at Boulder, Boulder, Colorado 80309-0390, USA}
\affiliation{$^{2}$Institute for Materials Research, Tohoku University, Sendai, Miyagi 980-8577, Japan}
\affiliation{$^{3}$Institute of Materials Structure Science, KEK, Oho, Tsukuba 305-0801, Japan}
\affiliation{$^{4}$Advanced Light Source, Lawrence Berkeley National Laboratory, Berkeley, California 94720, USA}

\date{\today}

\begin{abstract}
Electronic band dispersions in copper oxide superconductors have kinks around 70 meV that are typically attributed to coupling of electrons to a bosonic mode. We performed angle resolved photoemission spectroscopy (ARPES) experiments on overdoped cuprate high temperature superconductors to test the relationship between the superconducting transition temperature and electron-bosonic mode coupling. Remarkably, the kinks remain strong in the heavily overdoped region of the doping phase diagram of  La$_{2-x}$Sr$_x$CuO$_4$, even when the superconductivity completely disappears.  This unexpected observation is incompatible with the conventional picture of superconductivity mediated by the sharp bosonic modes that are responsible for the kink. Therefore, the pairing likely originates from something else, such as from interactions with a very broad electronic spectrum or from an unconventional mechanism without pairing glue.

\end{abstract}
\pacs{74.70.--b, 74.25.Jb, 79.60.--i} \maketitle

Conventional superconductivity arises from interactions between electrons and quantum atomic lattice vibrations (phonons): an electron can give off some of its energy and momentum to a phonon, which is then reabsorbed by a different electron. This effective interaction between electrons leads to the formation of Cooper pairs, which condense into the superconducting state. The superconducting transition temperature ($T_c$) is determined by the electronic density of states (DOS) near the Fermi energy, E$_F$, and the pairing function $\alpha^2F$ \cite{Carbotte}, which competes with pair-breaking interactions. Any other bosonic excitations such as magnetic fluctuations, charge fluctuations, plasmons, etc \cite{Abanov, Monthoux, Monthoux1} can replace phonons without significant modifications of the theory. The pairing function depends on the bosonic spectrum as well as on the strength of electron-boson coupling. Bosonic excitations that mediate pairing deform electronic dispersions near the Fermi surface, which can be measured by angle resolved photoemission (ARPES) \cite{ZXShen}. If the bosonic modes are sharp in energy, they should induce a kink in the dispersions from which $\alpha^2F$ may be extracted \cite{Grimvall}. 

High temperature superconductivity emerges when insulating perovskite copper oxides are doped and the Fermi surface appears. Discovery of the electronic dispersion kink near 70 meV by ARPES was interpreted as evidence that some of the excitations responsible for the kink are the pairing glue of high temperature superconductivity. \cite{Dahm, Johnston} Intense effort focused on uncovering the origin of the kink with phonons and magnetic fluctuations emerging as leading candidates \cite{Lanzara, Gromko, Kaminski, Johnson, TKKim, Cuk, Zhou1, Borisenko, Meevasana, Terashima, Graf, Park1, Iwasawa, Dahm}. Overdoped La$_{2-x}$Sr$_x$CuO$_4$ (LSCO) is well-suited for investigating pairing interactions in copper oxides, because this family of materials allows isolating the effect of pairing interactions from other factors that may affect superconductivity. Specifically, as we show below, the ARPES kink should disappear or become strongly suppressed when \textit{x} increases from 0.2 to 0.3 if it reflects the pairing interaction.

In this Letter, we report a surprisingly small reduction of the 70 meV kink magnitude with overdoping and, based on this observation, show that bosonic modes at this energy scale alone cannot mediate superconductivity. Thus either the conventional "pairing glue" picture is not valid or novel bosonic excitations whose spectra are at very different energies mediate superconductivity in the copper oxides.

We measured high quality single crystals of LSCO grown by the floating zone method. Measurements were performed on BL10 at the Advanced Light Source, which allows accurate sample alignment. Total energy and angular resolutions were 20 meV and 0.3$^{\circ}$, respectively. Sample temperature was 40 K, well above $T_c$ of LSCO (x=0.20) so that we can directly compare kink strengths for x=0.20 and x=0.30 without interference of the superconducting gap. Fermi surface (FS) mapping allowed us to align samples within 0.5$^{\circ}$ by making sure that the FS is symmetric (Fig. 1a,b). Our data were highly reproducible, which allowed us to reliably compare the strength of the ARPES kink in different samples.


\begin{figure}[h]
\centering \epsfxsize=8.7cm \epsfbox{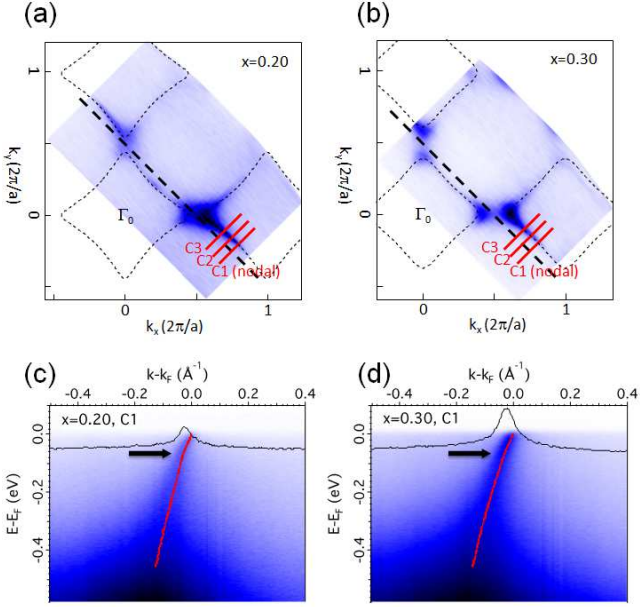}
\caption{Fermi surfaces and high energy band dispersions in LSCO. {\bf (a,b)} Constant energy mapping of ARPES intensity at E$_F$ integrating from -6 meV through 6 meV of LSCO {\bf (a)} (x=0.20) and {\bf (b)} (x=0.30). Straight dashed lines indicate the antiferromagnetic Brillouin zone boundary. Dotted lines indicate the FS extracted from a tight binding band model fit \cite{Yoshida2}. {\bf (c,d)} ARPES intensities along nodal direction (C1) of x=0.20 {\bf (c)} and x=0.30 {\bf (d)}. Red solid lines indicate band dispersions obtained by Lorentzian fitting of momentum distribution curves (MDCs). Arrows indicate the positions of ARPES kinks. Black solid lines represent MDCs at -0.05 eV, which are normalized by MDC peak area at -0.4 eV.
}
\label{Fig. 1}
\end{figure}


Figure 1 (a,b) shows that the Fermi surface (FS) in LSCO (x=0.20) may already be closed around $\Gamma$ point with the FS volume decreasing slightly at x=0.30 \cite{Yoshida2}. We performed self-energy analysis of the ARPES data along three red lines (C1, C2, C3) in 2nd Brillouin zone, (Fig. 1a,b) where the signal intensity is strongest and there is no significant matrix element variation \cite{Inosov}. Nodal band dispersions (Fig. 1c,d) follow a straight line from -0.5 eV to -0.1 eV. Clear deviation of the band dispersion away from a straight line near 70 meV (arrows) corresponds to the kink feature well known from previous studies. Current work focuses on this kink as well. 



\begin{figure}[h]
\centering \epsfxsize=8cm \epsfbox{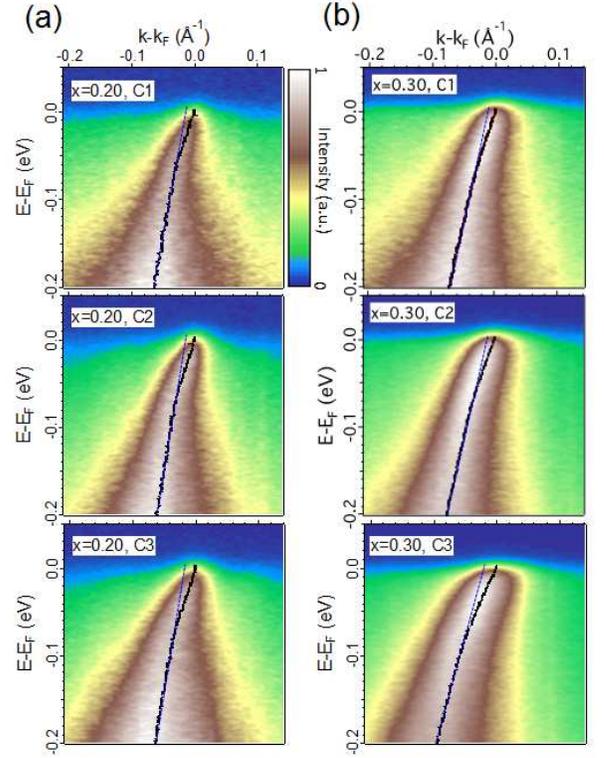}
\caption{ARPES intensity maps of {\bf (a)} LSCO (x=0.20) and {\bf (b)} LSCO (x=0.30) along C1 (nodal), C2 and C3 indicated in Fig. 1. Black solid lines on each map indicate band dispersions obtained by Lorentzian fitting of MDCs. Blue dotted straight lines fit to band dispersions at high binding energy and start to show deviation from black solid lines at about -0.07 eV.}
\label{Fig. 2}
\end{figure}

We zoom in on the ARPES data of LSCO (x=0.20, 0.30) to better see the kink in each cut (C1, C2, C3) in Fig. 2. Band dispersions clearly deviate from blue dotted straight lines starting at -0.07 eV regardless of doping concentrations and cuts. Kink strength of LSCO (x=0.30), surprisingly, seems to be as strong as that of LSCO (x=0.20), whereas $T_c$s of LSCO (x=0.20 and x=0.30) are 32 K and 0 K, respectively. 

There is no clear evidence for electron-bosonic mode coupling near the anti-node in LSCO (see Supplemental Material, Fig. S5 \cite{supp}). Whereas very strong electron-bosonic mode coupling signature at the anti-node in double or triple layer cuprates has been observed \cite{Kaminski, Gromko, TKKim, Cuk, Lee}, it is absent or very weak at the anti-node in single layer cuprates including not only LSCO but also Tl$_2$Ba$_2$CuO$_6$ and Bi$_2$Sr$_{1.6}$La$_{0.4}$CuO$_{6+\delta}$ \cite{Lee, Shi, Wei}. The strong contrast in electron-bosonic mode coupling between single and multi-layer cuprates may be due to either the difference in magnetic resonance modes \cite{Gromko, Johnson, Terashima, Shi} or almost no coupling between electrons and c-axis phonons (e.g. B$_{1g}$) in single layer cuprates because of symmetry \cite{Devereaux, Lee}. Thus three cuts (C1, C2, C3) represent a complete picture of electron-bosonic mode coupling in LSCO (x=0.20 and x=0.30).



\begin{figure}
\centering \epsfxsize=8.7cm \epsfbox{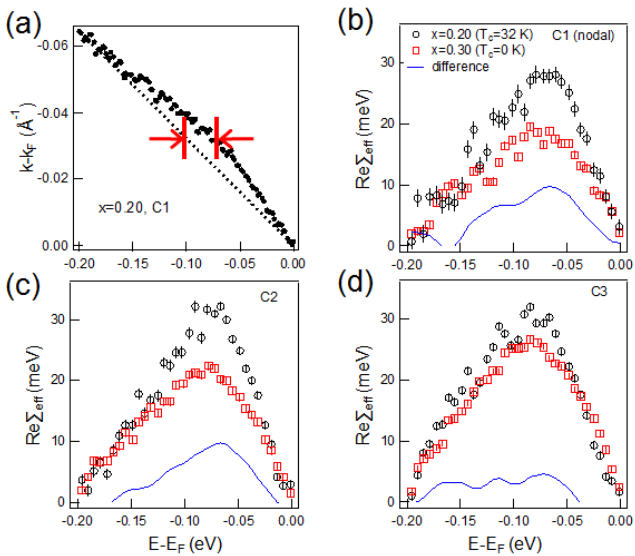}
\caption{ {\bf (a)} Band dispersion along the nodal direction extracted by lorenzian fits to MDCs. The dotted line indicates the linear bare band. Red arrows show Re$\Sigma_{eff}$ at - 0.07 eV. Circles and squares represent Re$\Sigma_{eff}$ of LSCO (x=0.20) and LSCO (x=0.30) respectively along C1 (nodal) {\bf (b)} , C2 {\bf (c)} and C3 {\bf (d)} directions, respectively. Blues lines represent the difference in Re$\Sigma_{eff}$ between x=0.2 and x=0.3.}
\label{Fig. 3}
\end{figure}

In order to estimate the strength of the 70 meV kink quantitatively, it is necessary to subtract the band dispersion in its absence  (bare band). The standard procedure followed in the vast majority of the literature is to assume a linear or, alternatively, a quadratic bare band, that meets the data at E$_F$ and 200 meV. Note that this ignores electron-electron interactions that make the bare band deviate from the measured band at 200 meV \cite{Kordyuk}. Since our conclusions do not depend on the functional form of the bare band (see below), we show only the results based on a linear bare band. The results based on a quadratic bare band are shown in Supplemental Material \cite{supp}.

Fig. 3 (b) shows the contribution of bosonic modes below 200 meV to the real part of the self energy (Re$\Sigma_{eff}$) extracted using the linear bare band connecting MDC peak positions at 0.2 eV and 0 eV binding energies (Fig. 3a) as is typical in such studies. Re$\Sigma_{eff}$ is much broader than expected from coupling only to a mode at 70 meV and indicates coupling to additional lower energy bosonic modes.  \cite{Meevasana}. Our Re$\Sigma_{eff}$, therefore, includes all bosonic modes below 200 meV. The kink strength is reduced from x=0.20 to x=0.30 by about 30$\%$ or less along C1, C2, and C3 directions (Fig. 3(b-d)), whereas $T_c$ changes from 32 K to 0. The kink at x=0.30 is similarly weaker than at x=0.20 in all cuts across the Fermi surface as shown in Fig. 3 (c,d). 



Previously Zhou \textit{et al}. investigated nodal electronic dispersion (along the 110-direction) in La$_{2-x}$Sr$_x$CuO$_4$ over the entire doping range \cite{Zhou2, Zhou3}. They reported universal Fermi velocity  as well as that the kink energy was independent of doping. It was also apparent in their published data that the kink strength dropped by almost an order of magnitude from x=0.22 to x=0.3, although the authors did not make this claim explicitly. Valla interpreted \cite{Valla} this feature as evidence for spin fluctuations, whose spectral weight goes down with overdoping \cite{Wakimoto, Lipscombe}. Our experiment aimed specifically at clarifying this issue shows that such a dramatic reduction does not occur. We presume that the difference between the previous results and ours is due to a better statistical quality of our results for x=0.3 and better sample alignment \cite{supp}.

To summarize, our experiments show that the kink magnitude does not drop significantly with overdoping even as $T_c$ goes to zero. We now present a simple argument that this result implies that excitations that induce the kink do not mediate superconductivity in the x=0.2 sample. 

In general, suppression of superconductivity upon changes in doping can occur due to the enhancement in pair-breaking, changes in the electronic structure, or the weakening of the pairing mechanism. Let's examine these possibilities one by one.

First, there is no solid evidence that increased doping enhances pair-breaking significantly enough to kill superconductivity. For example there is a suggestion that ferromagnetic fluctuations break Cooper pairs in the overdoped region \cite{Kopp}, but recent experiments do not support the presence of such fluctuations \cite{Kaiser}.

What about the electronic structure? Fig. 1 demonstrates that the FS topologies of the two samples are very similar. This result is consistent with previous systematic studies \cite{Yoshida}. Then, an important parameter determining $T_c$ is the electronic DOS at E$_F$. The FS volume of LSCO (x=0.30) is slightly smaller than that of LSCO (x=0.20) as shown in Fig. 1 (a,b), but other experimental results such as Hall measurement and optical conductivity show that the DOS at E$_F$ is bigger in LSCO (x=0.30) \cite{Uchida1, Takagi}. The discrepancy between results from ARPES and the other techniques can be understood by taking the quasiparticle spectral function near E$_F$ into account \cite{SAWATZKY, Trivedi}. Quasiparticle spectral function near E$_F$ is stronger in LSCO (x=0.30) as shown in Fig. 1(c,d), and the stronger quasiparticle spectral weight can be interpreted as higher DOS \cite{SAWATZKY, Trivedi}. Consequently, higher DOS at E$_F$ in LSCO (x=0.30) would tend to enhance superconductivity at x=0.3 vs. x=0.2, rather than suppress it.

Thus neither pair-breaking nor electronic structure evolution from x=0.2 to x=0.3 can suppress $T_c$, and  so within the conventional Eliashberg-type theories the only possible reason for the suppression of superconductivity is the reduction in the strength of the pairing interaction. Different ideas proposed for the pairing interaction in cuprates can be classified into the ones with a pairing glue and the ones without. All of these mechanisms rely on either electronic correlations due to the proximity to the Mott insulating state or collective excitations emerging from other order parameters, which may be related to the pseudogap \cite{Kivelson, Monthoux_review}. These effects should weaken on the overdoped side leading to the decrease in the tendency to form pairs. 

Different types of pairing glue have been proposed. For example, electron-electron interactions may induce bosonic collective modes, such as magnetic fluctuations, which then mediate superconductivity based on standard theory\cite{Dahm, Abanov}. Another possibility is that electronic correlations srongly enhance electron-phonon interactions \cite{Mishchenko}, which would allow phonons to mediate high $T_c$ superconductivity. More recently other types of pairing glue have been proposed such as charge density fluctuations \cite{Monthoux1} and current loop excitations \cite{Li, Varma1}. Our observation that the ARPES kink does not weaken strongly with overdoping allows us to rule out most of these possibilities.

All bosonic excitations that couple to conduction electrons add to the kink. Some meditate superconductivity (pairing glue) and others (non-pairing excitations) do not. For example, non-pairing excitations can be active in the nonsuperconducting s-wave channel and pairing glue can be active in the superconducting d-wave channel. The full spectrum of possibilities falls between two limiting cases described next. 

In the first case, the non-pairing component of the kink can be present in both samples, whereas pairing glue appears only in the superconducting x=0.2 sample. In this case the difference between Re$\Sigma_{eff}$ of the two samples (blue line in Fig. 3b-d) isolates the part of the kink that may originate from pairing glue. The maximum of this difference is about 10 meV, which corresponds to $T_c$ of no more than a few degrees in the case of YBa$_2$Cu$_3$O$_{7-x}$ \cite{Bohnen,Heid}. Considering no electron-bosonic mode coupling signature in LSCO at the anti-node, the estimated $T_c$ should be even lower in the case of LSCO.

On the other extreme, the entire kink may originate from the pairing glue, and the enhancement of the kink at x=0.2 may appear due to enhanced electron-boson coupling strength (for example due to reduced screening in the case of phonons \cite{Meevasana}) or due to the enhancement of the boson spectral weight. To explain the change in $T_c$ from 0 K to 32 K, superconductivity must appear only when $\alpha^2F$ exceeds a certain threshold. However, in this case, again, according to standard theory, the difference of 10 meV is not enough to raise $T_c$ from 0 K to 32 K (see Supplemental Material \cite{supp}). One can reach a similar conclusion by examining the cases intermediate between the two limiting cases.

We showed that if bosonic modes that serve as pairing glue are sharp in energy, they must induce pronounced electronic dispersion kinks, which must become strongly suppressed as $T_c$ drops with overdoping. Absence of this suppression means that such bosonic modes below 200 meV do not mediate high temperature superconductivity in the copper oxides. Specifically, this observation rules out both phonons and low energy magnetic fluctuations from the superconductivity mechanism. Phonons have sharp peaks in the DOS, which are expected to induce kinks. In fact we will argue in a different publications that the 70meV kink is consistent with an interaction with Cu-O bond-stretching phonons. Likewise low-energy magnetic fluctuations have features such as the 50 meV peak at optimal doping and the onset of the magnetic signal near 90meV in overdoped samples. These should also induce pronounced kinks if they are coupled to electronic quasiparticles strongly enough to mediate high $T_c$ superconductivity \cite{Dahm}. Thus we can use the same reasoning to rule out superconductivity mediated by low energy magnetic fluctuations. One caveat is that we cannot rule out bosonic modes below 20 meV such as an acoustic phonon with small momentum \cite{Plumb, Johnston2}, since our experimental resolution and statistics do not resolve kinks in such low energy scale. 

The pairing glue picture based on Eliashberg theory can survive only if the spectrum of excitations that mediates superconductivity is not expected to induce a kink in the electronic dispersion below 200meV. Evidence for such excitations comes from optical studies \cite{Heumen, Giannetti, Conte} and RIXS measurements \cite{Tacon}. STM experiments also point at high energy scales associated with superconductivity while indicating that features near 70 meV may be unrelated to pairing \cite{Pasupathy}. Another possibility is an entirely unconventional (non-BCS) mechanism of superconductivity that does not require pairing glue \cite{Anderson}. 

To summarize, we showed that doping dependence of the ARPES spectra of overdoped LSCO is inconsistent with pairing mediated by bosonic modes below 200 meV that are sharp in energy such as phonons or magnetic fluctuations. Instead our findings favor unconventional models of superconductivity based on pairing mediated by broad electronic spectra, high-energy excitations, or completely different mechanisms that do not require any pairing glue. 

\acknowledgements
The authors thank Ted Reber for valuable discussions. The Advanced Light Source is operated by the DOE, Office of Basic Energy Sciences. S.R.P. and D.R. were supported by the DOE, Office of Basic Energy Sciences, Office of Science, under Contract No. DE-SC0006939. Y.C., Q.W. and D.S.D. were supported by the DOE under Contract No. DE-FG02-03ER46066. The work at Tohoku University was supported by the Grant-In-Aid for Science Research A (22244039) from the MEXT of Japan.

\end{document}